# The Role of Physical Layer Security in Satellite-Based Networks

R. Singh, Ijaz Ahmad, Jyrki Huusko
VTT Technical Research Centre of Finland, Espoo, Finland

*Abstract*— **In the coming years, 6G will revolutionize the world with a large amount of bandwidth, high data rates, and extensive coverage in remote and rural areas. These goals can only be achieved by integrating terrestrial networks with non-terrestrial networks. On the other hand, these advancements are raising more concerns than other wireless links about malicious attacks on satellite-terrestrial links due to their openness. Over the years, physical layer security (PLS) has emerged as a good candidate to deal with security threats by exploring the randomness of wireless channels. In this direction, this paper reviews how PLS methods are implemented in satellite communications. Firstly, we discuss the ongoing research on satellite-based networks by highlighting the key points in the literature. Then, we revisit the research activities on PLS in satellite-based networks by categorizing the different system architectures. Finally, we highlight research directions and opportunities to leverage the PLS in future satellite-based networks.**

## I. INTRODUCTION

Over the past decade, significant achievements have been realized in wireless communication networking technologies. The next generation, 5G+, wireless systems will be deployed with standardized Internet of Things (IoT) techniques on a broad scale. It is widely accepted that the present and future wireless networks are expected to handle the incessant traffic demands of the transceiver pairs in the near future. For example, 5G and beyond (which is presently being researched) wireless networks can provide a tenfold higher throughput per end-user and drive the back-haul segment with a thousandfold more traffic. These technologies may be available to the market within the next two to three years. This is in line with ambitious goals set by European Commission (EU) in annual report 2019 to provide Gigabit connectivity to critical socio-economic drivers in the education and health sectors; and backbone networks of strategic importance [1]. The EU allocated EUR 3 billion to renew the Connecting Europe Facility [2]. The Finnish government also set similar pioneers for developing the digital infrastructure in Finland by 2025 to provide high-speed broadband access for all households and businesses [3]. The increasing number of mobile users day by day is further strengthening the need to yield better internet connectivity across all geographical locations. Despite of the advancements in terrestrial technologies, the deployment of terrestrial networks is limited due to the scarcity of infrastructure in difficult-to-serve areas such as remote rural areas, and some critical relevant application areas, such as remote or mobile healthcare.

On the other hand, it is reflected by standardization and industry efforts that the space communication technologies can be considered as an impetus for the forthcoming 5G networks and beyond due to its ability to provide broadband connectivity [4], [5]. The satellite links are capable of invigorating the available link capacity. They are also primed for both broadcast and multicast type services with higher throughput connections since they can serve a huge number of terrestrial users with a single transmission. The critical application of satellite communication is represented by the Digital Video Broadcasting (DVB) project (industry-led project supported by consortium of 270 broadcasters) in which DVB "second generation" (DVB-S2) standard and its extension (DVB-S2X) are currently being adopted [6]. Additionally, satellite communications can proficiently play a crucial role in providing emergency services during and after natural disasters.

Basically, the satellite networks comprise one or a number of satellites and are categorized into two different types depending on their orbit altitude, i.e., Geostationary (GEO) and Non-Geostationary (NGEO). The GEO satellites are capable of providing large coverage to cover one-third of the earth with an altitude of approximately 36,000 km, while NGEO satellites have relatively lower altitude and consist of several satellite constellations [7]. In this direction, tremendous efforts are devoted to develop Low Earth Orbit (LEO) constellations that can establish a transmission link with high-throughput and low latency. Previously, the development of multitude manufacturing and the launching of satellites was reserved only for governments (collaborating with international corporations). However, several private ventures have started developing satellite technologies with manufacturing and launch under the program named "*New Space*" in recent years. Several companies, including Amazon, SpaceX, TeleSAT, and OneWeb have already initiated big projects based on LEO satellites in which networks would be created using thousands of satellites. Moreover, a few of these companies have launched demo satellites. For instance, a Starlink constellation has been built by deploying 242 satellites by SpaceX by January 2020, which would be further extended to reach 42,000 satellites to develop a mega-constellation as a part of Starlink network [8]. In the same order, Google and Facebook launched their high-altitude platform (HAP) projects (Loon and Aquila, respectively) to provide internet access through balloons and drones [9], [10]. Although, both the projects have been defunct in 2021 and 2018, respectively, due to failure in long-term sustainability.

In the meantime, the research and academic community has initiated the third-generation partnership project (3GPP) as a study item for new radio (NR), i.e., 5G to cater to the upcoming requirements such as data offloading and service continuity [11]. In terrestrial mobile network (TMN), both the integrated satellite and stand-alone satellite systems can be supported by these technologies. The satellite links are advantageous over TMN in terms of latency, link stability, or burstiness. On comparing with TMN, the satellite connection has a very small round-trip time of up to 600-700 ms. This means that the high latency in satellite links lead to lower quality-of-experience (QoE) of end-user device, while broadband terrestrial links are responsible for poor QoE. On the other hand, satellite-terrestrial integrated network (STIN) is also considered as an appealing study item because of its increasing importance in the future 5G systems and beyond. The archetypal architecture of a STIN is depicted in Fig. 1. The recent advancement and exploitation of STINs support the fact that satellites can be proved as an effective means by providing necessary coverage to reach the areas beyond the terrestrial coverage to passengers in aircraft, trains, and vessels [12], [13]. Also, the satellite links are flexible as an integral part of the 5G ecosystem and can be the ideal candidate for high-reliability applications. However, there are some challenges in system design, such as higher signal delays and vulnerable links, etc., imposed due to the complicated communication environment of STIN. In Table I, previous studies with regard to space communication are summarized at a glance to highlight the contributions of these works.

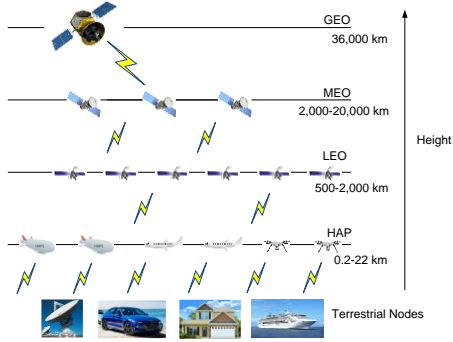

Fig. 1. Overview of satellite-based networks.

Table I
SUMMARY OF STUDIES ON SATELLITE NETWORKS

| Reference | Year | Key point |
|---|---|---|
| [29] | 2015 | An overview of secure handoff and secure transmission control |
| [30] | 2016 | A review of satellite communication systems for the Internet of Remote Things (IoRT). |
| [31] | 2016 | A survey on open system interconnection (OSI) model-based inter-satellite communications |
| [32] | 2016 | An overview of the current state-of-the-art of the satellite and terrestrial network convergence |
| [33] | 2017 | A survey on various challenges faced by free-space communication system in space communication |
| [34] | 2018 | An overview of survivability and scalability of satellite networks |
| [35] | 2018 | A review on software defined networking (SDN) based satellite networks |
| [36] | 2022 | An overview of security issues and solutions in satellite-terrestrial communications |
| [37] | 2022 | A survey on PLS and cryptographic approaches to deal security threats in satellite-based systems |

## II. PHYSICAL LAYER SECURITY

Despite all the advantages, the variety of IoT devices and machines in satellite networks are also susceptible to security threats due to broadcasting nature and wide coverage. This means that the free space wireless signals will not only be received by the intended user, but also any unauthorized user (which is located near to the intended user) could obtain the secure information due to power leakage in wireless signals. Therefore, immediate attention must be given to the privacy and security concerns in satellite networks. Traditionally, upper-layer encryption techniques are used to improve the confidentiality of satellite networks from eavesdroppers [14], [15]. However, encryption techniques have their own limitations when used in satellite communications. For example, the higher mobility of satellites results in higher bit error rate that makes it difficult for most encryption technologies. The higher mobility of satellite systems also makes the key management and sharing extremely challenging in the public key-based encryption technologies. Moreover, most IoT devices lack resources for fool-proof encryption technologies. Therefore, cryptographic techniques alone will not suffice in the realm of satellite communications, as discussed in [1]. Therefore, other approaches such as physical layer security (PLS), which was introduced recently in satellite networks [16], must be adopted.

On the other hand, in PLS, the protection against eavesdropping attacks is provided by exploiting the random nature of wireless channels (e.g., fading, noise, and interference) without relying on the private key as long as the quality of the legitimate link gets worse than the illegitimate link [17]-[22]. Unlike cryptographic schemes, the PLS do not require highly computing abilities at the communication entities and not involves any upper-layer protocols with complex calculations.

This makes the reception of secret information at the unauthorized user difficult, and if received, then instigate incorrect decoding of the received information, which enhances the security of the overall system. It is worth mentioning that recent advancements in beamforming techniques and signal processing with multi-input and multi-output (MIMO) methods are having significantly favorable impacts on security and privacy of the system [23]. The way of PLS applicability in the development of wireless systems is paved due to its irrefragable advantages. In the last few years, extensive literature regarding secure transmission in the context of satellite communications has been developed using the framework of PLS with a focus on recent designs and solutions. Earlier, the focus of the researchers was on the investigation of security of land mobile satellite systems (LMSs) using an information-theoretic PLS approach. Afterward, the research interests were shifted to secrecy analysis of hybrid satellite-terrestrial relay networks (HSTRNs). Provoked by the recent technological progress in space information networks, information security issues in STINs have already been addressed using PLS, see [24]-[28] and references therein. As a summary, the earlier works on PLS in space communication are provided in Tables II, III, and IV. The studies based on LMS, HSTRN, and STIN are compared to present an overview of the advancements in PLS approaches in satellite-terrestrial networks.

TABLE II
SUMMARY OF STUDIES ON LMS COMMUNICATION NETWORKS

| Target problems | PLS Approach | Assessment/Performance Metrics | Types of satellite | Link Type | CSI (Known or unknown) | Number of illegitimate receivers | Reference | Year |
|---|---|---|---|---|---|---|---|---|
| Confidentiality | Information-theoretic | Simulations/ Secrecy outage probability | Spot-beam | Downlink | Known | Single | [38] | 2016 |
| Confidentiality | Information-theoretic | Simulations/ Secrecy outage probability | Spot-beam | Downlink | Both | Multiple | [39] | 2018 |
| Confidentiality | Information-theoretic | Simulations/ Secrecy outage probability | Spot-beam | Downlink | Known | Multiple | [40] | 2019 |
| Confidentiality | XOR Network Coding | Simulations/ Average secrecy rate | Multi-beam | Downlink and Uplink | Known | Two | [41] | 2015 |
| Confidentiality | Information-theoretic | Theoretical/ Secrecy capacity | Multi-beam | Downlink | Known | Multiple | [42] | 2017 |
| Confidentiality | Spoofing detection | Simulations/ Bit error rate | Dual-polarized | Downlink | Known | Single | [43] | 2017 |
| Confidentiality | Polarization filtering | Simulations/ Bit error rate | Dual-polarized | Downlink | Known (imperfect) | Single | [44] | 2017 |
| Confidentiality | Artificial Noise | Simulations/ Secrecy capacity | Satellite-relay | Uplink | Known | Single | [45] | 2018 |
| Confidentiality | Information-theoretic | Simulations/ Secrecy outage probability | Spot-beam | Downlink | Known (imperfect) | Single | [46] | 2019 |
| Confidentiality | Beamforming | Simulations/ Average secrecy rate | Multi-beam | Downlink | Known (imperfect) | Single | [47] | 2019 |
| Confidentiality | Information-theoretic | Simulations/ Secrecy capacity | Spot-beam | Downlink | Known | Multiple | [48] | 2019 |
| Confidentiality | Information-theoretic | Simulations/ Sacrifice rate | Spot-beam | Downlink | Known | Single | [49] | 2019 |
| Confidentiality | Beamforming | Simulations/ Secrecy energy efficiency | Multi-beam | Downlink | Known (imperfect) | Multiple | [50] | 2019 |
| Authentication | Doppler frequency shift based-authentication | Simulations/ False alarm rate and miss detection rate | Spot-beam | Downlink | Known | Single | [51] | 2021 |
| Authentication | Deep learning | Simulations/ Average Area Under the Curve | Multi-beam | Downlink | Both | Single | [52] | 2023 |

TABLE III
SUMMARY OF STUDIES WORK ON HSTRNS

| Target problems | PLS Approach | Assessment/Performance Metrics | Link Type/System Setup | Number of relays and protocol | CSI (Known or unknown) | Number of illegitimate receivers | Reference | Year |
|---|---|---|---|---|---|---|---|---|
| Confidentiality | Artificial Noise with paired carrier multiple access | Simulations/ Secrecy capacity | Downlink/SIMO | Single (AF) | Known | Multiple | [53] | 2018 |
| Confidentiality | Information-theoretic | Simulations/ Secrecy outage probability | Downlink/SIMO | Multiple (AF) | Known | Multiple | [54] | 2017 |
| Confidentiality | Information-theoretic | Simulations/ Secrecy outage probability | Downlink/SIMO | Multiple (DF) | Known | Single | [55] | 2017 |
| Confidentiality | Information-theoretic | Simulations/ Secrecy outage probability | Downlink/SIMO | Multiple (DF) | Unknown | Single | [56] | 2019 |
| Confidentiality | Information-theoretic | Simulations/ Secrecy capacity | Downlink/SIMO | Multiple (DF) | Known | Multiple | [57] | 2019 |
| Confidentiality | Information-theoretic | Simulations/ Secrecy outage probability | Downlink/SIMO | Multiple (Both AF and DF) | Known | Multiple | [58] | 2019 |
| Confidentiality | Artificial Noise | Simulations/ Bit error rate | Downlink/SIMO | Multiple (AF) | Known | Single | [59] | 2018 |
| Confidentiality | Artificial Noise | Simulations/ Outage probability | Downlink/SIMO | Multiple (DF) | Known | Single | [60] | 2021 |
| Confidentiality | Information-theoretic | Simulations/ Secrecy outage probability | Downlink/SIMO | Multiple (DF) | Both | Multiple | [61] | 2022 |

TABLE IV
SUMMARY OF STUDIES ON STINS

| Target problems | PLS Approach | Assessment/Performance Metrics | Link Type/System Setup | Number of terrestrial mobile networks | CSI (Known or unknown) | Number of illegitimate receivers | Reference | Year |
|---|---|---|---|---|---|---|---|---|
| Confidentiality | Successive convex approximation | Simulations/ Transmit power | Downlink/SIMO | Single | Known | Multiple | [62] | 2019 |
| Confidentiality | Penalty dual decomposition | Simulations/ Achievable secrecy rate | Downlink/SISO | Single | Known | Single | [63] | 2018 |
| Confidentiality | Beamforming with artificial noise | Simulations/ Total transmit power | Downlink/SIMO | Multiple | Known | Multiple | [64] | 2018 |
| Confidentiality | Beamforming | Simulations/ Achievable secrecy rate | Downlink/SIMO | Single | Known | Multiple | [65] | 2018 |
| Confidentiality | Information-theoretic | Simulations/ Secrecy outage probability | Downlink/SIMO | Multiple | Known | Multiple | [66] | 2020 |
| Confidentiality | Beamforming | Simulations/ Secrecy capacity | Downlink/SIMO | Multiple | Known | Multiple | [67] | 2022 |

## III. FUTURE RESEARCH DIRECTIONS

Several attempts have been consecrated in the literature to design and analyze secure communications in satellite networks (including backhaul satellite networks and satellite-terrestrial networks). It is the inescapable fact that satellites have potential applications in the next-generation 5G technologies, but the research in space communication is still in its genesis, and many issues of immediate attention are in need of further investigation. In the following, a few of the interesting research items are mentioned to cope with challenges in the emerging network architecture, which have not been covered in the previous studies.

### A. Intelligent Eavesdropping

Recently, a few studies pointed towards the idea of intelligent eavesdropping in which eavesdroppers have high signal processing capabilities, antenna resources, and self-sustainable (i.e., energy harvesting) power sources [68]. These eavesdroppers can be designed in such a way that they are simultaneously capable of wiretapping secure information and contaminating pilot signals if operated in full-duplex mode. In this scenario, the intelligent eavesdropper can easily extract the full CSI of the indented ground receiver and disturb the CSI sharing between the satellite and the receiver [69]. Moreover, it will be practicable for the intelligent eavesdropper to nullify the applied security procedure such as friendly jamming, transceiver diversity, and secure beamforming. Towards this end, the security problems of such satellite-based systems in which upgraded eavesdroppers are placed in the transmission path is a promising direction for future research that must be addressed.

### B. Outdated CSI, imperfect CSI and unavailability of CSI

The outdated and imperfect CSI significantly affects the secrecy performance evaluation. The imperfect CSI is caused due to error in channel estimation, whereas users' mobility may cause outdated CSI. On the other hand, in the passive eavesdropping scenario, the adversary tries to hide itself from the legitimate transmitter and receiver and never shares its CSI with them. In the absence of CSI (or with outdated and imperfect CSI), the transmitter can never estimate the ergodic capacity of the illegitimate link, and hence perfect secrecy is compromised. In the case of a satellite-based system, the round-trip propagation delay creates a high latency and result in difficulty in estimating the exact CSI. On the other side, if the user is moving faster than the CSI update speed, the CSI acquired by the satellite may become outdated. It is noticeable that no study has evaluated the secrecy performance of the satellite-based system with outdated CSI. Towards this end, the secrecy analysis of such systems will be a promising direction for future work to elaborate on the impacts of the uncertainty availability of CSI.

### C. Multi-user information-theoretic security

Another interesting research item may be physical layer secrecy for the practical systems where multiple users can transmit and receive simultaneously (two-way or full-duplex communication). In recent years, the MIMO techniques showed its potential in enhancing system performance. Therefore, we will exploit the MIMO techniques in future work. This is motivated by the recent study [70] on PLS in the presence of multiple eavesdroppers. However, the researchers have limited their analysis to one-way communication only. The virtues of MIMO technology further motivate us to extend the idea of MIMO in space communication.

### D. Cross-Layer Security Mechanism

By 2025, more than 75 billion different IoT devices are connected everywhere and sharing their data with each other. In the case of space communication, the registers are used to share or access information or datasets. Therefore, a strong security mechanism must be built to defend the information or data against both active and passive eavesdropping attacks as the framework is open to the public. It is the requirement for the next-generation satellite-based systems to design new robust security schemes which exploit the advantages of both the upper-layer cryptographic and physical-layer approaches [71]. By adopting a hybrid security technique, the next level of security can be achieved in future satellite-terrestrial systems.

### E. Unmanned aerial vehicle (UAV)-enabled satellite-terrestrial network

The UAVs are the potential solution to improve the network capacity and increase the coverage range at a very low cost as they can serve as flying mobile base stations [72]. The UAVs are designed in such a way so that they can move freely in the sky to provide a large coverage. The demands of efficient and comprehensive services for the future 6G can be met by integrating satellite, terrestrial, and airborne networks together to develop a cooperative network by utilizing multimodal and multidimensional information [73]. However, the existing literature still considers satellite, terrestrial, and airborne as three different networks. Furthermore, an increasing number of UAVs may cause security concerns as UAVs can also be used to eavesdrop the secure information. Therefore, it is of most concern to design a three-tier heterogeneous network that comprises dynamic topology and complex structure and is capable of cover deep space, near-earth space, airborne, and ground.

*F. FSO-based satellite communication*

With daily up gradation in technology and fast-growing demand for high data rates, wireless communication often faces scarcity of RF spectrum and reaching out its limits. Hence, the replenishment of spectrum scarcity has become a primary research concern in the wireless communication domain. In an attempt to overcome the spectrum crunch and provide high data throughput, FSO technology has been introduced as a possible solution for 5G and beyond cellular networks with a cost-efficient framework [74]. In satellite communication, the RF links can be replaced by FSO links to reduce the cost of the entire system as well as to enhance security and reliability. Recently, few projects such as loon by Google and Aquila by Facebook were launched to test the feasibility of FSO transmissions at high altitude platforms. A few studies also investigated the system performance for satellite-terrestrial downlink systems where the link between the satellite and terrestrial node was assumed as an FSO link.

*G. Satellite-UAV-Terrestrial Underwater Communications*

The potential applications of underwater wireless communication (UWC) in environmental monitoring, offshore exploration, disaster precaution, and military operations have increased information security concerns [75]. Like terrestrial wireless communication, the information-theoretic approach can be used to enhance the security and reliability of the UWC systems as the data in UWC is transmitted through wireless carriers in the unguided water environment. However, research on the secrecy performance of UWC systems is still in its infancy despite the growing number of UWC applications. Future research can be extended in designing UAV-assisted satellite-terrestrial underwater systems where UAVs can serve as relays for underwater sensor nodes and can also be capable of tracking the position of the nodes.

*H. Artificial Intelligence (AI) enabled PLS*

In information-theoretic approach utilizes the stochastic nature of wireless channels, such as fading and noise, to secure the information without generating any secret key. For better secrecy, it is necessary to estimate the physical layer imperfections perfectly. In recent years, artificial intelligence (AI) techniques for channel estimation have become the research community's focus. Based on the estimated CSI, the base station can adopt the secrecy rate to conceal the information from eavesdroppers. Moreover, several machine learning (ML) techniques, including supervised and unsupervised learning, can be used for physical layer authentication in a heterogeneous environment [76]. Until now, the AI-enabled PLS has not been explored in the satellite-based system, which can be a critical research problem to be investigated in the near future.

IV. CONCLUSION

With the advent of 6G in the future, we are aiming for high-speed, widespread connectivity with low latency. These objectives can only be achieved by exploiting the inherent characteristics of satellite-based networks. On the contrary, the security threats on satellite-terrestrial links are raising concerns as these links are more vulnerable due to the openness of the channels and heterogeneity of the connected devices. The PLS has proved its ability to handle the security issues in the existing network architecture and has the potential to compensate for the security risks in future networks. In this paper, we provide a thorough examination of the current state-of-the-art in the field of PLS in satellite communication. We have also presented an overview of various satellite network architectures. Finally, this paper provides future research directions by highlighting several open research issues.

ACKNOWLEDGMENT

This work has been supported in part by Business Finland through the SUNSET-6G and the AI-NET-ANTILLAS projects.